\documentclass[conference]{IEEEtran}
\IEEEoverridecommandlockouts
\usepackage{cite}
\usepackage{amsmath,amssymb,amsfonts}
\usepackage{algorithmic}
\usepackage{graphicx}
\usepackage{textcomp}
\usepackage{xcolor}
\usepackage{soul}
\usepackage{booktabs}
\usepackage[left=0.680in,right=0.680in,top=0.76in,bottom=1.05in]{geometry}
\def\BibTeX{{\rm B\kern-.05em{\sc i\kern-.025em b}\kern-.08em
    T\kern-.1667em\lower.7ex\hbox{E}\kern-.125emX}}

\begin{document}

\title{Towards In-Cabin Monitoring: A Preliminary Study on Sensors Data Collection and Analysis
\thanks{This work has been carried out in cooperation with Sleep Advice Technologies S.r.l. (\url{https://www.satechnology.eu/}) and it has been financially supported by the UE project  A-IQ READY that receives funding within the Key Digital Technologies Joint Undertaking (KDT JU) - the Public-Private Partnership for research, development and innovation under Horizon Europe – and National Authorities under grant agreement n. 101096658.}
}

\author{\IEEEauthorblockN{Jacopo Sini, Luigi Pugliese, Sara Groppo, Michele Guagnano, Massimo Violante\\
\textit{Politecnico di Torino}\\
Turin, Italy \\
\{jacopo.sini, luigi.pugliese, sara.groppo, michele.guagnano, massimo.violante\}@polito.it}}


\maketitle

\begin{abstract}
The last decade's market has been characterized by wearable devices, mainly smartwatches, edge, and cloud computing. A possible application of these technologies is to improve the safety of dangerous activities, especially driving motor vehicles. Common enabling technologies, such as system-on-chip, ultra-low-power computational platforms, and wide-band wireless connectivity, push all these trends. On the one hand, wearable devices, thanks to the continuous contact with the user's body, can measure physiological parameters. On the other hand, edge computing and machine learning techniques, alongside cameras, allow the implementation of contactless computer vision systems capable of providing information about the user's current behavior. Another trend is the usage of RADARs in automotive applications, both for collision avoidance and monitoring driver behavior. These technologies can be combined to develop systems designed to aid the driver. For the sake of this paper, we are focusing on warning drivers, allowing them to know whenever they are drowsy and hence risking a sleep onset or are not paying attention to the road. Developing such systems poses many challenges, such as automatic classification of physiological signal patterns, facial expression recognition, head movements and eye gaze detection. These challenges have been individually addressed in the literature. Anyway, we noticed a need for more description on implementing data fusion. Two main reasons for adopting the fusion approach are to improve the quality of the overall representation (increasing accuracy and specificity against drowsy) and make a more reliable system due to redundancy.
\end{abstract}


\section{Introduction}
As reported by American National Highway Traffic Safety Administration (NHTSA), driver loss of attention is a worldwide problem that leads to a high number of deaths every year. The leading cause of the decrease in attention while driving is drowsiness. Particularly, sleep at the wheel contributes to motor vehicle accidents for 20\% of all police-reported crashes\cite{b1}. 
Several methods for monitoring the driver's state have been developed, mainly focusing on the way to drive \cite{b2}, the camera-based measurements\cite{b4}, and the physiological parameters of the driver \cite{b7}. 
Detecting methods based on the way to drive have low reliability because obtained results can change a lot due to unpredictable factors such as road geometry and traffic conditions.
Camera-based measurements are theoretically very effective in drowsiness detection but, in real driving conditions, they are subjected to light and skin color variations \cite{b8}. 
Physiological parameters that can be used for drowsiness study are electrocardiography (ECG), electromyography (EMG), electrooculography (EOG), electroencephalography (EEG), and respiration rate (RR) \cite{b9}. They can provide very accurate results but some of them, such as EEG, require very invasive sensors to be acquired, and this makes it difficult to use them for driver drowsiness detection. The most used physiological parameters are heart rate (HR), heart rate variability (HRV), and respiration rate (RR).
The best approach for driver monitoring is combining data from multiple sensors, obtaining data that are more consistent and reliable \cite{b26}.
This process is known as sensor-data fusion and its great reliability is nowadays seen as a way to increase road safety \cite{b10}.
Our research group is involved in studying sleep macro and micro patterns, sleep disorders, and daytime drowsiness related to sleep onset while driving.
In this study, data from different sensors have been collected and paired as preliminary work for a fusion algorithm able to give a comprehensive analysis of camera-based and physiological-based parameters of the driver to evaluate drowsiness and attention levels. A low attention level is of course dangerous by itself and could be compromised by drowsiness and/or attention to the road, and these two factors are crucial for motor vehicle accidents.
A vital sign detection RADAR was used to measure HR and RR, a smartwatch was used to detect HR, RR, and HRV, and two cameras were used to find eye blinking (EB), eye gaze angles (EGA), percentage of eyelid closure over the pupil over time (PERCLOS) and head movements (HM).

\section{State of the Art}

Regarding physiological parameters, the most accountable for sleep onset prediction are:
\begin{itemize}
    \item Heart Rate (HR), which describes the contractions of the heart per minute.
    \item Heart Rate Variability (HRV), which represents the change in time intervals between adjacent heartbeats.
    \item Respiration Rate (RR), which is the number of breaths a person takes per minute.
\end{itemize}
These physiological parameters are the most reliable and accurate in drowsiness detection as they are concerned with what is happening with the driver physically; in fact, through their monitoring, it is possible to advise the driver to stop the vehicle before the physical symptoms of drowsiness effectively appear \cite{b24}. Moreover, they can be extracted both through contact or contactless technologies\cite{b16}.\\
Concerning contact-based solutions, heart signals such as electrocardiogram (ECG) and photoplethysmography (PPG) are considered accurate measures of fatigue. However, their use is limited because of the intrusive nature of the sensors. Nevertheless, novel sensors can be embedded in the steering wheel or the seat belt \cite{b60}. As proposed by Jung et al., electrodes were embedded in the steering wheel; in this way, HR and HRV were extracted from ECG and drowsiness was successfully detected. However, very highly accurate sensors were needed and the position of the hands of the driver on the steering wheel was crucial for the proper acquisition of the data\cite{b61}. In another study, Li and Chuang developed a PPG sensor placed on the steering wheel of the vehicle. In this case, HRV was extracted from the raw PPG signal. Then, a Support Vector Machine (SVM) was trained to classify the state of the driver as fatigued or alert, thus obtaining a 95\% fatigue detection accuracy. Even considering the non-intrusiveness of this method, it is susceptible to human error and natural movements\cite{b62}. 
Recently, wrist-worn wearable devices have been successfully employed as contact-based solutions for monitoring physiological parameters \cite{b25}. As reported by Kundinger et al., by the use of various smartwatch devices an accuracy of about 92\% was reached in detecting drowsiness, compared to medical-grade device\cite{b63}. However, the main limitation of these devices can be the 
accuracy of sensing technology; in addition, it is required that the subject wears the device, relying on the diligence of the driver itself.
Considering the challenges of contact-based methods, various contactless solutions have been recently explored for monitoring physiological parameters. Among them, techniques based on radar have shown promising results. Liu et al. developed a novel radar-based technology for drowsiness detection, where the accuracy of heart rate was 96.4\% and the accuracy of drowsiness detection was 82.9\%, compared to other state-of-the-art approaches\cite{b64}. 
Moreover, camera-based techniques offer the possibility to estimate drowsiness through eyes and head tracking.
Eye gaze direction is widely used to detect drowsiness.
In-Ho Choi and Yong-Guk Kim \cite{b29} track the driver's gaze direction by tracing the pupil's center point.  
Eye blinking 
becomes slower and more frequent while getting drowsier \cite{b30}.
Shekari Soleimanloo S et al. \cite{b32} used Optalert to detect eye blinking, an infrared oculography system.
Wang X and Xu C used instead the smarteye eye tracker \cite{b33}.
PERCLOS express for how much time there is at least 70$\%$ (or 80$\%$) of eyelid closed during a unit of time of 1 minute (or 30 seconds).
In \cite{b35}, B. Bakker et al. used a 3-camera smart eye pro system, with infrared lighting. 
G. Du et al. used instead an RGB camera for PERCLOS detection \cite{b36}. 
Head movements are another visual drowsiness sign as, while falling asleep, someone could start nodding. 
They can be found by recording with a camera \cite{b37},
or by using a magnetic tracker, as the Ascension flock of birds \cite{b38}. 

\section{Proposed methodology}
The system has been designed to obtain a time series of physiological parameters from the previously described sensors.
Each sensor communicates with the experimenter workstation with different protocols and physical interfaces.
For this reason, we decided to design an application developed for the Microsoft Windows environment to save a CSV file containing the logs from the current experimental campaign, aligned from the timing point of view, and to send, as a JSON record via MQTT, each acquisition to a centralized NoSQL database.

Message Queuing Telemetry Transport (MQTT) allows the creation of a client-server system. In this protocol, the server (in charge of relaying the record to the clients) is called \emph{broker}.  A client can have two roles: if it receives the relayed data, it is called \emph{subscriber} while, if it instead sends records, it is a \emph{publisher}. For this reason, MQTT is defined as a publish-subscribe protocol.

JavaScript Object Notation (JSON) is a data-exchange format. It allows to create a hierarchical data structure, and it is widely used in web development due to its convenience: thanks to ready-to-use serializer and deserializer functions, it is possible respectively to encapsulate data contained in a class into a JSON string or vice-versa to extract the JSON string content creating novel instances of the contained classes.

The designed system is capable of retrieving  data from the following devices:
\begin{itemize}
 \item Vital Sign Detection Radar (Chuhang Technology Radar), featuring HR, RR, and distance measurement. The latter value is needed to assess the quality of the provided data. If the distance changes it means that there is a movement of the person with respect to the Radar and hence that the measure is unreliable. The Radar is directly connected via USB (emulated serial port, with data formatted in binary form) with the experimenter workstation.
 \item Wearable Device (GARMIN Enduro, Enduro 2, VenuSq, Venu 2) featuring off-the-shelf HR, RR, HRV, and a drowsiness onset real-time prediction algorithm developed by our research group. Bluetooth wireless protocol connects the Device with an Android smartphone, which relays the data to a remote MQTT broker that transmits a copy to its subscribers and stores the received data into a persistence layer implemented resorting to MongoDB.
 \item Camera (Varroc and one developed by our research team) featuring EB, EGA, PERCLOS, and HM.
 \end{itemize}

As it is possible to observe, some data of these acquisitions are redundant (HR and RR), but it is important to keep both measurements due to the different measurement technologies. In particular, the RR is more reliable when measured by the radar, while the HR is more reliable when measured by the wearable device.

\begin{figure*}[!h]
\centering
\includegraphics[width=15cm]{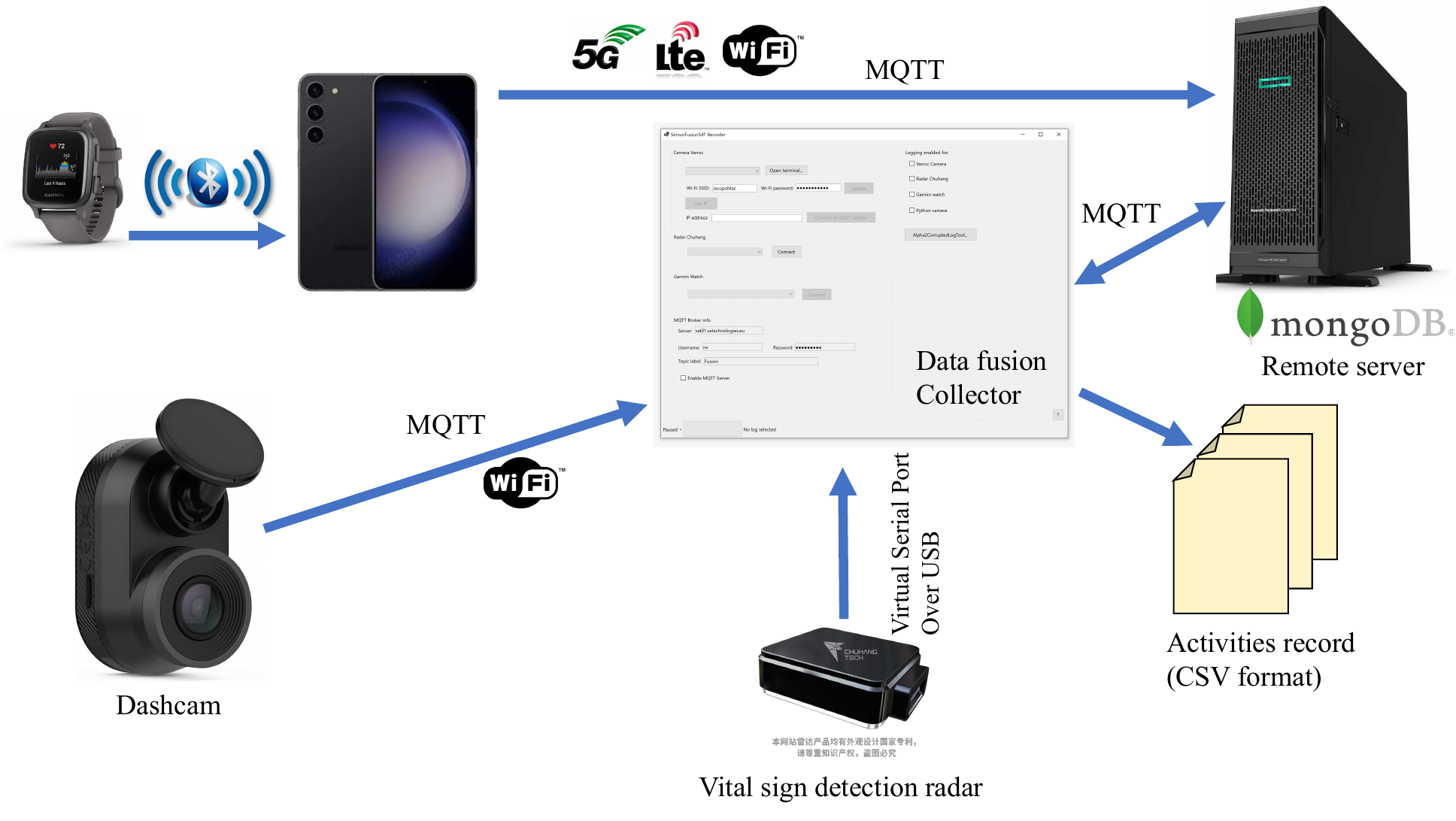} 
\caption{The proposed architecture to perform Data Fusion.}
\label{arch_figure}
\end{figure*}

In Fig.\ref{arch_figure} it is shown the architecture of the system. The central node is the workstation running the Data fusion collector. It receives data from the dashcam (via a Wi-Fi or Ethernet connection), resorting to the MQTT protocol, and from the vital sign detection radar, via a virtual serial port over USB. The data from the wearable device are received indirectly: the smartwatch sends the data to an app running on a smartphone via a virtual serial port over Bluetooth. The smartphone sends the data via Internet (mobile communication standards or Wi-Fi), resorting to the MQTT protocol to a remote server, which runs a Broker that receives the data and send them back to the Data Fusion Collector (in this case, acting as a Subscriber). Finally, the Data fusion collector saves the received data into a CSV file in the workstation's local file system and sends it to the remote server via MQTT (in this case, acting as a Publisher).
The remote server stores all the received data in mongoDB.
The data transmitted over MQTT are encapsulated, resorting to the JSON format, allowing to store into the mongoDB directly the received records.

The variety of devices involved justifies the choice of a NoSQL database: it is possible to use different dashcams, wearable devices, and vital sign detection radar, each with a different set of recorded measurements. Thanks to the NoSQL database (with does not feature fixed-structure tables) and JSON format, it is possible to change the record structure seamlessly.

Currently, the system is a work in progress, so we expect to add more sensors to this set in the following months, mainly to measure environmental conditions like air relative humidity and temperature, ambient light sensors, and air quality parameters, in particular, carbon dioxide quantity. 

Another future expansion of this system will be the possibility of generating synchronized time series in offline mode: the devices are equipped with proprietary tools to connect them to the workstation, which can create log files: for using these for data fusion, it is needed to align all the data from the timing point of view: the application will read the logs and generate a newer one with the fusion. Moreover, it can also store in the NoSQL database the data fusion to make data available for remote access and long-term reliable storage.

The purpose of the system data-fusion system is twofold: 
\begin{enumerate}
    \item \label{phase1} in this phase, to aid the experiments described in Section \ref{experimental-results}, needed for the development and testing of the data fusion algorithm.
    \item \label{phase2}since the data fusion collector has the availability of all the data from the sensors acts as a proof-of-concept of the final system running the real-time warning system.
\end{enumerate}

The data collected in phase \ref{phase1} are currently used to develop the real-time algorithm iteratively: the algorithm runs offline, and the results are compared with respect to the sleep onset detected from the various polysomnography read by sleep expert medical doctors. 

Considering a possible future use in real vehicles (phase \ref{phase2}), the architectural choices may appear clearer: the dashboard/infotainment computer system can run the data fusion algorithm and warn the driver. It simplifies the management of the camera and radar: they can be part of the car itself, directly connected to the car power supply, and wired to the onboard systems. This direct connection allows for fulfilling the mandatory need for driver drowsiness and attention warning (DDAW) (Regulation (EU) 2019/2144). In any case, when the driver wears a smartwatch, connecting their smartphone to the system is more convenient by considering the typical commercial use: the driver connects their phone via Bluetooth to the car's infotainment system. Exploiting this link to seamlessly exchange data about drowsiness to the data fusion algorithm through the infotainment system is possible.
Looking forward, proposing a communication standard to allow different vendors to implement this functionality in their products appears to be a good idea.

\section{Experimental results}
\label{experimental-results}
In this first stage, we conducted some experiments (Mantainance Weakfullness Tests) involving 16 volunteers, 5 females, 11 males, age range 25-30 to verify the sensitivity and accuracy of the data fusion algorithm. 

Each test lasted about 1.5 h, considering the instrumentation with the polysomnograph and its pads and wirings and the set-up of the data fusion system.
After the instrumentation phase, the volunteers are asked to take two test sessions of 20 minutes of Mantainance Weakfullness Test (MWT). Between the two tests, the volunteers are required to answer two questionnaires. The first questionnaire, compiled before the tests, aimed to understand the participants' clinical status and his/her capability to stay awake (ESS). The second questionnaire used the Karolinska Sleepiness Scale to evaluate their drowsiness level during the test. 

The fusion approach has undergone thorough evaluation to validate its effectiveness. This evaluation encompasses two key aspects: firstly, enhancing the overall representation's quality, including the information about distractions, and secondly, bolstering the system's reliability through redundancy.
\begin{figure}[!h]
    \centering
    \includegraphics[width=9cm]{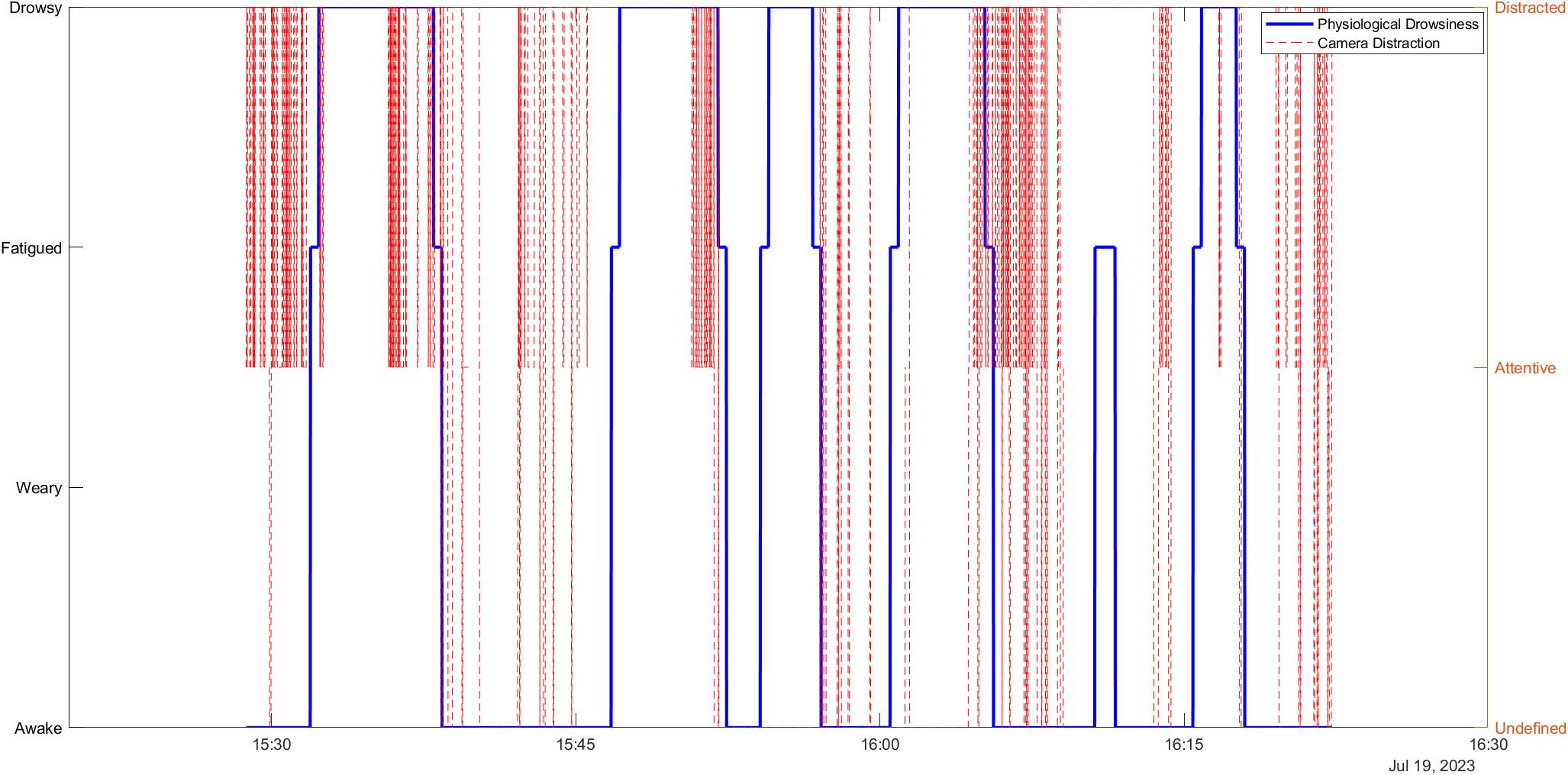}
    \caption{Comparison between physiological-based drowsiness detection and camera-based attention detection. On the vertical axis on the left (black) are reported the outcomes from the wearable device, while on the axis on the right (orange) are the outcomes from the camera.}
    \label{Attention}
\end{figure}
In Fig.\ref{Attention}, it can be observed that a complete understanding of the passenger's status requires additional information about distraction, which is not directly evident from the physiological data.
\begin{figure}[!h]
    \centering
    \includegraphics[width=9cm]{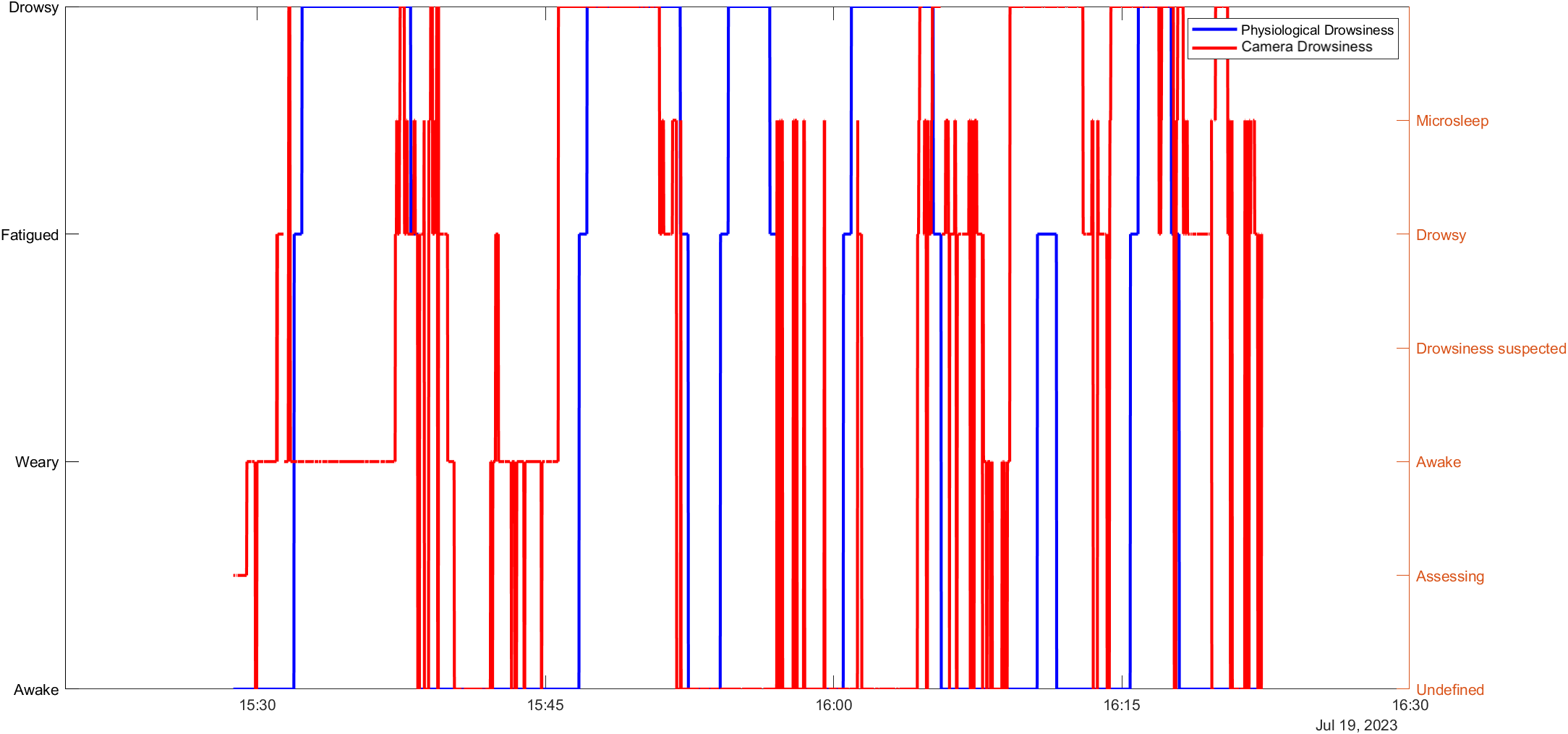}
    \caption{Comparison between physiological-based (left axis) and camera-based (right axis) detected drowsiness level.}
    \label{Drowsy}
\end{figure}
In Fig.\ref{Drowsy}, it is evident that the physiological-based method provides clearer and smoother information regarding the drowsiness level. Notably, the blue line indicates potential alarms for the passenger minutes earlier than the red line. To further enhance the system's performance, implementing system redundancy to establish a clearer initial driver condition would be beneficial.
\section{Conclusions}
In this study, a significant array of sensors was utilized, each capable of operating independently and improving performance by leveraging information from other devices.

The results section examined a specific real-life scenario involving a combination of a camera and a wearable device. This integrated system provided a comprehensive understanding of the driver's status in terms of both drowsiness and attention levels.

Currently, the investigation is focused on merging information from the various sensors utilized. The future direction of this research involves the development of a real-time sensor fusion algorithm to effectively synthesize data from all sensors simultaneously. This advancement aims to further streamline and enhance the overall system's performance.

The algorithm development is ongoing as the radar results have not been presented yet due to lower data accuracy, specifically concerning the wearable data.

\vspace{12pt}


\begin{thebibliography}{00}


\bibitem{b1} WHO Road Traffic Injuries Report, \url{https://www.who.int/news-room/fact-sheets/detail/road-traffic-injuries}, last visited 2023/04/26.

\bibitem{b2} G. Zhenhai, L. DinhDat, H. Hongyu, Y. Ziwen and W. Xinyu, "Driver Drowsiness Detection Based on Time Series Analysis of Steering Wheel Angular Velocity," 2017 9th International Conference on Measuring Technology and Mechatronics Automation (ICMTMA), Changsha, China, 2017, pp. 99-101, doi: 10.1109/ICMTMA.2017.0031.


\bibitem{b4} O. Khunpisuth, T. Chotchinasri, V. Koschakosai and N. Hnoohom, "Driver Drowsiness Detection Using Eye-Closeness Detection," 2016 12th International Conference on Signal-Image Technology \& Internet-Based Systems (SITIS), Naples, Italy, 2016, pp. 661-668, doi: 10.1109/SITIS.2016.110.



\bibitem{b7} L. Pugliese, M. Violante and S. Groppo, "A Novel Algorithm for Detecting the Drowsiness Onset in Real-Time," in IEEE Access, vol. 10, pp. 42601-42606, 2022, doi: 10.1109/ACCESS.2022.3167708.

\bibitem{b8} Sahayadhas A, Sundaraj K, Murugappan M. Detecting driver drowsiness based on sensors: a review. Sensors (Basel). 2012 Dec 7;12(12):16937-53. doi: 10.3390/s121216937. Erratum in: Sensors (Basel). 2021 Jan 11;21(2): PMID: 23223151; PMCID: PMC3571819.

\bibitem{b9} Siddiqui HUR, Saleem AA, Brown R, Bademci B, Lee E, Rustam F, Dudley S. Non-Invasive Driver Drowsiness Detection System. Sensors (Basel). 2021 Jul 15;21(14):4833. doi: 10.3390/s21144833. PMID: 34300572; PMCID: PMC8309856.

\bibitem{b26} Marceddu, A.C.; Pugliese, L.; Sini, J.; Espinosa, G.R.; Amel Solouki, M.; Chiavassa, P.; Giusto, E.; Montrucchio, B.; Violante, M.; De Pace, F. A Novel Redundant Validation IoT System for Affective Learning Based on Facial Expressions and Biological Signals. Sensors 2022, 22, 2773. https://doi.org/10.3390/s22072773 

\bibitem{b10} Deo A, Palade V, Huda MN. Centralised and Decentralised Sensor Fusion-Based Emergency Brake Assist. Sensors (Basel). 2021 Aug 11;21(16):5422. doi: 10.3390/s21165422. PMID: 34450863; PMCID: PMC8400951.

\bibitem{b24} L. Pugliese, M. Violante and S. Groppo, "Real-time sleep prediction using a virtual sensor to estimate Heart Rate Variability through Respiratory Rate," 2022 IEEE 16th International Conference on Application of Information and Communication Technologies (AICT), Washington DC, DC, USA, 2022, pp. 1-4, doi: 10.1109/AICT55583.2022.10013549.

\bibitem{b16} M. Ramzan, H. U. Khan, S. M. Awan, A. Ismail, M. Ilyas, and A. Mahmood, “A Survey on State-of-the-Art Drowsiness Detection Techniques,” IEEE Access, vol. 7, pp. 61904–61919, 2019, doi: 10.1109/ACCESS.2019.2914373.

\bibitem{b60} Sikander, G., Anwar, S. (2019). Driver Fatigue Detection Systems: A Review. IEEE Transactions on Intelligent Transportation Systems, 2339–2352. https://doi.org/10.1109/TITS.2018.2868499

\bibitem{b61} Jung, S. J., Shin, H. S., Chung, W. Y. (2014). Driver fatigue and drowsiness monitoring system with embedded electrocardiogram sensor on steering wheel. IET Intelligent Transport Systems, 43–50. https://doi.org/10.1049/iet-its.2012.

\bibitem{b62} Li, G., Chung, W. Y. (2013). Detection of driver drowsiness using wavelet analysis of heart rate variability and a support vector machine classifier. Sensors (Switzerland), 16494–16511. https://doi.org/10.3390/s131216494

\bibitem{b25} (In press) L. Pugliese, M. Violante and R. Groppo, "Real-time Sleep Prediction Algorithm using Commercial Off the Shelf Wearable Devices," 2023 IEEE Smart World Congress, Porstmouth, UK, 2023, pp. 1-4.

\bibitem{b63} Kundinger, T., Sofra, N., Riener, A. (2020). Assessment of the potential of wrist-worn wearable sensors for driver drowsiness detection. Sensors (Switzerland). https://doi.org/10.3390/s20041029

\bibitem{b64}Liu, S., Zhao, L., Yang, X., Du, Y., Li, M., Zhu, X., Dai, Z. (2022). Remote Drowsiness Detection Based on the mmWave FMCW Radar. IEEE Sensors Journal, 15222–15234. https://doi.org/10.1109/JSEN.2022.3186486
















\bibitem{b29} In-Ho Choi and Yong-Guk Kim, "Head pose and gaze direction tracking for detecting a drowsy driver," 2014 International Conference on Big Data and Smart Computing (BIGCOMP), Bangkok, 2014, pp. 241-244, doi: 10.1109/BIGCOMP.2014.6741444.

\bibitem{b30} B. G. Pratama, I. Ardiyanto and T. B. Adji, "A review on driver drowsiness based on image, bio-signal, and driver behavior," 2017 3rd International Conference on Science and Technology - Computer (ICST), Yogyakarta, Indonesia, 2017, pp. 70-75, doi: 10.1109/ICSTC.2017.8011855.


\bibitem{b32} Shekari Soleimanloo S, Wilkinson VE, Cori JM, Westlake J, Stevens B, Downey LA, Shiferaw BA, Rajaratnam SMW, Howard ME. Eye-Blink Parameters Detect On-Road Track-Driving Impairment Following Severe Sleep Deprivation. J Clin Sleep Med. 2019 Sep 15;15(9):1271-1284. doi: 10.5664/jcsm.7918. PMID: 31538598; PMCID: PMC6760410.

\bibitem{b33} Wang X, Xu C. Driver drowsiness detection based on non-intrusive metrics considering individual specifics. Accid Anal Prev. 2016 Oct;95(Pt B):350-357. doi: 10.1016/j.aap.2015.09.002. Epub 2015 Oct 1. PMID: 26433567.


\bibitem{b35} B. Bakker et al., "A Multi-Stage, Multi-Feature Machine Learning Approach to Detect Driver Sleepiness in Naturalistic Road Driving Conditions," in IEEE Transactions on Intelligent Transportation Systems, vol. 23, no. 5, pp. 4791-4800, May 2022, doi: 10.1109/TITS.2021.3090272.

\bibitem{b36} G. Du, L. Zhang, K. Su, X. Wang, S. Teng and P. X. Liu, "A Multimodal Fusion Fatigue Driving Detection Method Based on Heart Rate and PERCLOS," in IEEE Transactions on Intelligent Transportation Systems, vol. 23, no. 11, pp. 21810-21820, Nov. 2022, doi: 10.1109/TITS.2022.3176973.

\bibitem{b37} R. Oyini Mbouna, S. G. Kong and M. -G. Chun, "Visual Analysis of Eye State and Head Pose for Driver Alertness Monitoring," in IEEE Transactions on Intelligent Transportation Systems, vol. 14, no. 3, pp. 1462-1469, Sept. 2013, doi: 10.1109/TITS.2013.2262098.

\bibitem{b38} J. C. Popieul, P. Simon and P. Loslever, "Using driver's head movements evolution as a drowsiness indicator," IEEE IV2003 Intelligent Vehicles Symposium. Proceedings (Cat. No.03TH8683), Columbus, OH, USA, 2003, pp. 616-621, doi: 10.1109/IVS.2003.1212983.









\end{thebibliography}
\end{document}